\documentclass[11pt]{article}
\usepackage{hyperref}
\pdfoutput=1
\begin{document}
\title{The bounce-splash of a viscoelastic drop}
\author{Federico Hern\'andez-S\'anchez, Ren\'e Ledesma \& Roberto Zenit \\
\\\vspace{6pt} Instituto de Investigaciones en Materiales \\ Universidad Nacional Aut\'onoma de M\'exico \\
Cd. Universitaria, M\'exico D.F., M\'EXICO} \maketitle
\begin{abstract}
This is an entry for the Gallery of Fluid Motion of the 61st
Annual Meeting of the APS-DFD ( fluid dynamics videos ). This
video shows the collision and rebound of viscoelastic drops
against a solid wall. Using a high speed camera, the process of
approach, contact and rebound of drops of a viscoelastic liquid is
observed. We found that these drops first splash, similar to what
is observed in Newtonian colliding drops; after a few instants,
the liquid recoils, recovering its original drop shape and bounce
off the wall.
\end{abstract}

\section{Introduction}

We are interested in studying non-Newtonian phenomena in general;
in particular, we have studied the motion of spheres and bubbles
in complex fluids\cite{soto2006,{mendoza2008}}. We also have
previously studied the collision of solid spheres immersed in
viscous liquids\cite{joseph2001} and more recently the collision
of bubbles\cite{zenit2008}. This visualization project shows the
continuation of our previous work in these two subjects.

\section{Experimental Conditions}
We release viscoelastic drops of approximately 2.2 cm in diameter
from different heights in air to produce collisions at various
speeds against a solid surface. Different aqueous mixtures of
gelatin were used as the test liquid. By varying the viscoelastic
properties of the drops and the collision speed we can change the
value of the relevant dimensionless number. The Weissemberg number
is defined as
\begin{equation}\label{We}
    We=\frac{\lambda U}{D}
\end{equation}
where $\lambda$ is the relaxation time, $U$ is the impact velocity
and $D$ is the drop diameter. Hence, a large value of $We$ would
show a liquid-like behavior; on the other hand, a small $We$
collision would resemble a solid rebound. In this  video
contribution the collision process of drops for a range of values
of Weissemberg numbers is shown.

The most interesting case shown is that for the largest $We$. At
contact, the drops splashes: the whole liquid volume spreads into
a thin sheet. After a few instants, as a result of the elasticity
of the media, the liquid sheet recoils. The drop is re-formed, but
highly distorted. Nevertheless, the drop bounces off the wall and
exhibits large shape oscillations.

To our knowledge this process has not been studied in the past.

\section{Videos}

Our video contributions can be found at:

\begin{itemize}
    \item \href{http://hdl.handle.net/1813/11474}{Video 1, mpeg2, full
resolution}

    \item \href{http://hdl.handle.net/1813/11476}{Video 2, mpeg1, low
resolution}

\end{itemize}

\end{document}